\documentclass[graybox]{SNmult}

\usepackage{type1cm}        
\usepackage{makeidx}         
\usepackage{graphicx}        
\usepackage{multicol}        
\usepackage[bottom]{footmisc}

\usepackage{newtxtext}       %
\usepackage[varvw]{newtxmath}       

\makeindex             

\begin{document}
\title*{Renormalization-group perspective on spontaneous stochasticity}
\author{Alexei A. Mailybaev 
and Luca Moriconi
}
\institute{Alexei A. Mailybaev \at IMPA, Rio de Janeiro, Brazil, \email{alexei@impa.br}
\and Luca Moriconi \at IF-UFRJ, Rio de Janeiro, Brazil, \email{moriconi@if.ufrj.br}}

\maketitle
\abstract*{We present a renormalization-group perspective on spontaneous stochasticity in hydrodynamic turbulence, viewed through the lens of multiscale dynamical systems. Building on previously established results for a solvable multiscale Arnold’s cat model, we show that spontaneous stochasticity emerges as a universal fixed point of an RG transformation acting on Markov kernels, independent of the microscopic regularization. Classical examples—including the Feigenbaum equation, the central limit theorem, and hierarchical spin models—are reinterpreted within the same framework, placing spontaneous stochasticity alongside other universality phenomena.
}

\abstract{We present a renormalization-group perspective on spontaneous stochasticity in hydrodynamic turbulence, viewed through the lens of multiscale dynamical systems. Building on previously established results for a solvable multiscale Arnold’s cat model, we show that spontaneous stochasticity emerges as a universal fixed point of an RG transformation acting on Markov kernels, independent of the microscopic regularization. Classical examples—including the Feigenbaum equation, the central limit theorem, and hierarchical spin models—are reinterpreted within the same framework, placing spontaneous stochasticity alongside other universality phenomena.
}

\section{Introduction}

The problem of prediction in developed turbulence remains one of the central
challenges in physics and applied mathematics~\cite{frisch1999turbulence}.
In many practical applications---including atmospheric and oceanic flows,
as well as engineering turbulence modeling---one seeks to predict
large-scale observables in systems characterized by extremely wide ranges
of dynamically interacting scales.
A classical line of research, initiated by Lorenz in 1969~\cite{lorenz1969predictability}
and further developed in
\cite{leith1972predictability,ruelle1979microscopic,eyink1996turbulence,palmer2014real,boffetta2017chaos},
emphasized the role of scale-local interactions in the rapid amplification of
small perturbations, which severely limits predictability.
These works highlight a fundamental feature of turbulence:
microscopic fluctuations can propagate to macroscopic scales over finite time.

Such observations are closely related to the phenomenon of
\emph{spontaneous stochasticity}, whereby the ideal (zero-noise) limit of a turbulent system remains intrinsically stochastic~\cite{eyink2025beyond},
even though the limiting system is formally deterministic.
Spontaneous stochasticity has been extensively studied in the Lagrangian
framework, where non-uniqueness of particle trajectories in rough velocity
fields leads to stochastic behavior in deterministic
limits~\cite{bernard1998slow,falkovich2001particles,eyink2020renormalization,drivas2024statistical,barlet2025spontaneous}.
It also admits an Eulerian formulation~\cite{mailybaev2016spontaneously,thalabard2020butterfly,bandak2024spontaneous},
where non-uniqueness arises at the level of fields or weak solutions~\cite{eyink2024space}.
In the present work we focus exclusively on the Eulerian perspective,
formulating spontaneous stochasticity as a probabilistic resolution of
non-uniqueness in multiscale dynamical systems.

Renormalization-group (RG) ideas have long played an important role in
turbulence theory.
Classical RG approaches, inspired by the Kadanoff--Wilson framework~\cite{wilson1983renormalization,cardy1996scaling},
aim to derive effective large-scale dynamics through coarse-graining and
rescaling procedures~\cite{forster1977large,yakhot1986renormalization,goldenfeld2018lectures,inverseRG,canet2022functional}.
More recently, RG techniques have been applied directly to spontaneous
stochasticity~\cite{mailybaev2023spontaneous,mailybaev2025rg,mailybaev2025sabra}.
However, the resulting RG formulation differs structurally from the
Kadanoff--Wilson scheme and instead resembles functional RG equations of the
Feigenbaum type~\cite{feigenbaum1983universal}, namely, an iteration in the functional space defined through composition and rescaling.

The primary goal of this work is to place the RG formulation of spontaneous
stochasticity within the broader landscape of renormalization-group theories.
We show that the RG relation arising in the multiscale Arnold's cat model~\cite{mailybaev2023spontaneously},
previously introduced as a solvable example of Eulerian spontaneous
stochasticity, can be viewed alongside classical RG constructions such as
the Feigenbaum functional equation~\cite{feigenbaum1983universal}, the central limit theorem~\cite{sinai1992probability}, and
hierarchical models in statistical physics~\cite{dyson1969existence}.
These examples can be unified by interpreting them as different realizations
of dynamics defined on the same self-similar (fractal) space--time lattice,
with distinct local dynamical rules specifying the RG operator.

The paper is organized as follows.
In Section~\ref{sec2} we recall the phenomenology of noise propagation in turbulence and the concept of spontaneous stochasticity.
In Section~\ref{sec3} we consider the multiscale Arnold's cat model as a solvable example exhibiting spontaneous stochasticity.
In Section~\ref{sec4} we formulate the associated RG relation in terms of flow maps and
Markov kernels and analyze the resulting RG fixed point.
Subsequently, we reinterpret several classical RG constructions---including
the Feigenbaum equation (Section~\ref{sec5}), the central limit theorem (Section~\ref{sec6}), and hierarchical spin
models (Section~\ref{sec7})---within the same lattice framework.
Finally, we discuss the conceptual implications of this unifying viewpoint
for universality and for the role of spontaneous stochasticity in
turbulence theory.

\section{Spontaneous stochasticity: the physical picture}
\label{sec2}

A defining feature of turbulence is its extreme sensitivity to small-scale effects. Even minute perturbations—far below observational or numerical resolution—can rapidly influence the large-scale dynamics. Understanding how such microscopic influences survive, or even dominate, macroscopic behavior is one of the central conceptual challenges of turbulence theory. In this section, we introduce the notion of \emph{spontaneous stochasticity}, which provides a natural framework for addressing this problem and sets the stage for a renormalization-group (RG) interpretation discussed later in the paper.

\subsection{Fluctuating hydrodynamics} 

A convenient starting point is the incompressible Navier--Stokes equations supplemented with microscopic fluctuations, as originally proposed in the Landau--Lifshitz theory of fluctuating hydrodynamics. In dimensionless variables, these equations read \cite{landau1959fluid,bandak2022dissipation}
\begin{equation}
\label{eq1_1}
\partial_t \mathbf{u} + \mathbf{u}\cdot\nabla \mathbf{u}
= -\nabla p + \mathrm{Re}^{-1}\Delta \mathbf{u}
+ \sqrt{\Theta}\,\nabla\cdot\boldsymbol{\xi} + \mathbf{f},
\quad \nabla\cdot\mathbf{u}=0 .
\end{equation}
In the three-dimensional setting, $\mathbf{u}(\mathbf{x},t) \in \mathbb{R}^3$ is the velocity field depending on time $t$ in a triply periodic domain $\mathbf{x}\in\mathbb{T}^3$, $p(\mathbf{x},t) \in \mathbb{R}$ is the pressure field enforcing incompressibility, and $\mathbf{f}(\mathbf{x},t) \in \mathbb{R}^3$ is a deterministic forcing acting at large scales. The random $3\times 3$ tensor field $\boldsymbol{\xi}(\mathbf{x},t)$ represents microscopic (thermal) fluctuations and is modeled as Gaussian white noise with a covariance chosen to preserve incompressibility and isotropy, i.e. $\langle \xi_{ij}(\mathbf{x},t) \xi_{kl}(\mathbf{x}',t') \rangle = \left( \delta_{ik}\delta_{jl}+\delta_{il}\delta_{jk}-\frac{2}{3}\delta_{ij}\delta_{kl} \right) \delta^3(\mathbf{x}-\mathbf{x}')\delta(t-t')$.

Equation~\eqref{eq1_1} depends on two dimensionless control parameters. The Reynolds number $\mathrm{Re}=UL/\nu$ measures the importance of inertia (characteristic velocity $U$ and length $L$) relative to viscosity $\nu$. The dimensionless noise parameter $\Theta = 2\nu k_B T/(\rho L^4 U^3)$
encodes the strength of microscopic thermal fluctuations through the
temperature $T$, Boltzmann constant $k_B$, and mass density $\rho$. In addition, the continuum description itself breaks down below a molecular length scale $\ell_{\mathrm{mic}}$. One can represent this by introducing a Galerkin cutoff in the Fourier space. This cutoff reduces the system to a finite (but very large) set of stochastic differential equations corresponding to wave numbers $|\mathbf{k}|\le 2\pi/\delta$, where $\delta=\ell_{\mathrm{mic}}/L$, ensuring mathematical well-posedness while preserving the multiscale structure of the flow \cite{bandak2022dissipation,eyink2024space}.

We focus on the initial value problem with deterministic initial data
\begin{equation}
\label{eq1_1IC}
\mathbf{u}(\mathbf{x},0)=\mathbf{u}_0(\mathbf{x}).
\end{equation}
Because of the noise term in~\eqref{eq1_1}, the solution for $t>0$ is a stochastic process describing a probability distribution over velocity fields. Importantly, randomness enters the dynamics only through microscopic fluctuations: the initial condition and the large-scale forcing are fixed and deterministic.

\subsection{Ideal limit and spontaneous stochasticity}

In physically relevant turbulent flows, all three parameters—$\mathrm{Re}^{-1}$, $\Theta$, and $\delta$—are extraordinarily small. For example, in the atmospheric boundary layer one finds the order-of-magnitude estimates \cite{garratt1994atmospheric,bandak2024spontaneous}
\begin{equation}
\label{eq1_3}
\mathrm{Re}^{-1}\sim 10^{-8},\quad
\Theta\sim 10^{-38},\quad
\delta\sim 10^{-10}.
\end{equation}
These values highlight the enormous separation between the macroscopic flow scales and the microscopic scales where dissipation, noise, and molecular structure become relevant.

The standard theoretical approach is to remove these small effects sequentially: one first neglects noise and microscopic cutoffs, recovering the deterministic Navier--Stokes equations, and then takes the inviscid limit $\mathrm{Re}\to\infty$, formally obtaining the Euler equations for ideal fluid:
\begin{equation}
\label{eq1_1Euler}
\partial_t \mathbf{u} + \mathbf{u}\cdot\nabla \mathbf{u}
= -\nabla p + \mathbf{f},
\quad \nabla\cdot\mathbf{u}=0 .
\end{equation}
However, existing numerical and theoretical evidence indicates that
time-dependent solutions of the Navier--Stokes equations do not converge,
in general, as $\mathrm{Re}\to\infty$, but may only admit convergence,
if at all, along subsequences.
This failure of convergence signals that the order in which limits are taken matters, and that some information carried by microscopic effects may survive at large scales.

The idea of \emph{spontaneous stochasticity} is to reformulate the idealization procedure itself. Rather than taking the small parameters to zero one by one, one considers an \emph{ideal limit} in which viscosity, noise, and microscopic cutoffs vanish simultaneously \cite{bandak2024spontaneous,ortiz2025spontaneous}. A representative scaling is
\begin{equation}
\label{eq1_4}
\mathrm{Re}^{-1}\to 0,\quad
\Theta=\mathrm{Re}^{-\beta}\to 0,\quad
\delta=\mathrm{Re}^{-\gamma}\to 0,
\end{equation}
with suitable exponents $\beta$ and $\gamma$, consistent with~\eqref{eq1_3}
for $\beta = 19/4$ and $\gamma = 5/4$.
This formulation may be viewed either as the result of a physically consistent parameter tuning~\cite{bandak2024spontaneous}, or more abstractly
as encoding that all microscopic effects are asymptotically small and
interrelated through scale separation.

Formally, the limit~\eqref{eq1_4} still leads to the Euler equations~\eqref{eq1_1Euler}. The key question is whether the \emph{solutions} of the fluctuating system~\eqref{eq1_1} converge, and if so, what kind of object they converge to. The spontaneous stochasticity hypothesis asserts that the limit exists, but is not deterministic. Instead, one expects the following properties \cite{mailybaev2015stochastic,mailybaev2016spontaneously,thalabard2020butterfly,bandak2024spontaneous,ortiz2025spontaneous}:

\emph{(i) Convergence:}
The solutions of~\eqref{eq1_1} with initial condition~\eqref{eq1_1IC} converge in distribution in the ideal limit;

\emph{(ii) Stochasticity:}  The limiting dynamics is genuinely random,
i.e., defines a stochastic process in time;

\emph{(iii) Spontaneity:} Individual realizations of the limiting process solve the deterministic Euler equations~\eqref{eq1_1Euler} in a weak sense;

\emph{(iv) Universality:} The limiting stochastic process does not depend on the detailed manner in which the ideal limit is taken.

Taken together, these properties describe a striking scenario: although the Euler equations are formally deterministic, their physically relevant solutions are selected probabilistically. A necessary ingredient for such behavior is the non-uniqueness of weak solutions to the Euler equations \cite{de2010admissibility,daneri2021non}, but non-uniqueness alone does not explain why a particular probability law should emerge, nor why it should be universal.

Universality is the most conceptually demanding aspect of spontaneous stochasticity. It suggests that large-scale turbulent statistics are insensitive to microscopic modeling details—such as the precise form of dissipation, noise, or cutoff—and depend only on the scale-invariant structure of the ideal (Euler) dynamics. This perspective naturally calls for a renormalization-group formalism,
in which the ideal limit is understood as a fixed point (or attractor)
of the RG flow.

At present, a comprehensive understanding of spontaneous stochasticity
for the full three-dimensional Navier--Stokes equations remains out of reach.
Accordingly, we regard the above hypothesis as a research program
rather than a settled statement.
Nevertheless, simplified models provide compelling and increasingly
quantitative support for this scenario.
In particular, shell models of turbulence offer a clean framework
in which spontaneous stochasticity and universality can be demonstrated
and analyzed in detail
\cite{mailybaev2016spontaneously,bandak2024spontaneous,mailybaev2025sabra}.
For the Navier--Stokes equations themselves, numerical studies have reported
convergence toward stochastic solutions in two-dimensional flows generated
by vortex-sheet initial data, closely related to the
Kelvin--Helmholtz instability
\cite{fjordholm2016computation,thalabard2020butterfly}.
More recently, spontaneous stochasticity has also been established
in Navier--Stokes models posed on logarithmic lattices,
providing a bridge between continuum dynamics and multiscale toy models
\cite{ortiz2025spontaneous}.

\subsection{Phenomenology of noise dynamics}

\begin{figure}[tp]
\centering
\includegraphics[width=0.95\textwidth]{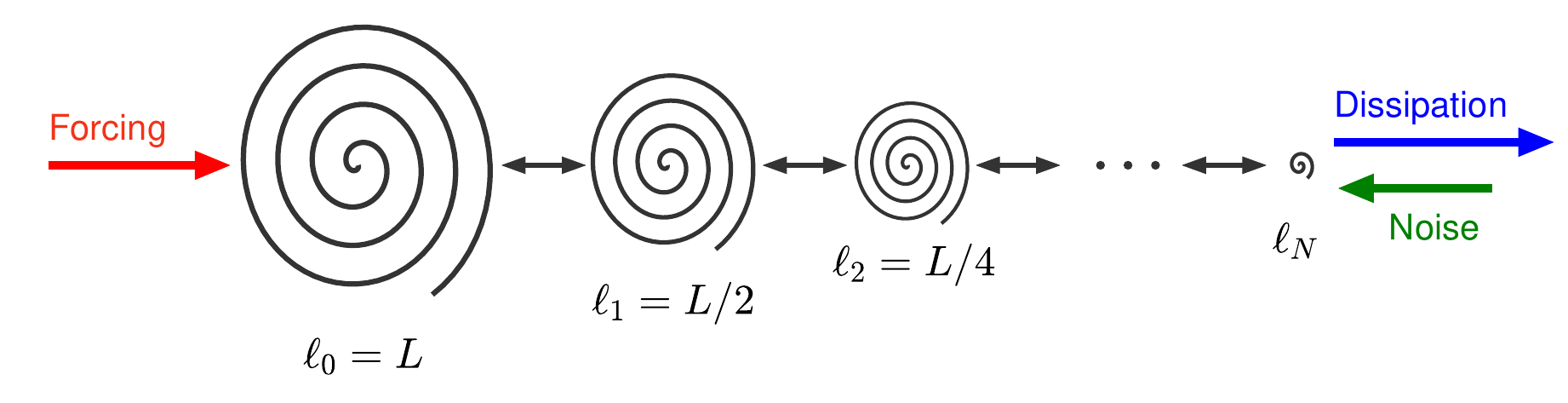}
\caption{Schematic description of turbulence: energy is injected at the largest
scale $L$ and transported through a hierarchy of progressively smaller eddies
toward the dissipation scale, while microscopic noise propagates in the opposite
direction, from small to large scales.}
\label{fig1}
\end{figure}

A turbulent flow is commonly described as a hierarchy of interacting eddies,
spanning a wide range of spatial scales~\cite{frisch1999turbulence}.
Energy is injected at large scales and subsequently transferred through a
cascade of progressively smaller eddies until it is ultimately dissipated at
small scales; see Fig.~\ref{fig1}.
The range of scales between the forcing and dissipation scales is referred to as
the inertial interval.
Within this phenomenological picture, eddies of size $\ell$ are characterized by
a typical velocity increment $\delta u(\ell)$ and an associated turnover time
$\tau(\ell)\sim \ell/\delta u(\ell)$.
In the inertial interval, both the spatial scales and the corresponding turnover
times decrease as power laws, and interactions are assumed to be local in scale:
eddies at a given size interact most strongly with eddies of comparable size, or
with their immediate neighbors in the cascade.

An important implication of this multiscale structure concerns the propagation
of perturbations across scales~\cite{lorenz1969predictability,leith1972predictability,eyink1996turbulence}.
Consider a fluctuation introduced at the smallest dynamically active
scale of the flow, namely, the dissipation scale.
This fluctuation interacts with slightly larger eddies over a time of order
given by the corresponding turnover time, then propagates to the next larger
scale, and so on.
The total time required for information to travel from the smallest to the
largest scale is therefore given by a sum of turnover times across the cascade.
Since these times form a geometric sequence, this sum converges to a finite
value.

To make this argument concrete, consider a dyadic hierarchy of scales
\begin{equation}
\ell_n = 2^{-n}L,\quad n=0,1,2,\dots,
\end{equation}
where $L$ denotes the integral (forcing) scale.
Assume that velocity increments obey a H\"older scaling law,
$\delta u(\ell)\sim U(\ell/L)^h$, where $U$ is a characteristic large-scale
velocity and $h\in(0,1)$ is the H\"older exponent (with $h=1/3$ in Kolmogorov's
K41 theory).
The corresponding turnover time at scale $\ell_n$ is then
\begin{equation}
\tau_n \sim \frac{\ell_n}{\delta u(\ell_n)}
\sim 2^{(h-1)n}\tau_0, \quad \tau_0 = \frac{L}{U}.
\end{equation}
Consider now the propagation of a fluctuation from the smallest active scale
$\ell_N$ up to the largest scale $L=\ell_0$ through successive interactions with
neighboring scales.
The total propagation time can be estimated as the sum of turnover times along
the cascade,
\begin{equation}
T_{0\leftarrow N}\ \sim\ \sum_{n=0}^{N}\tau_n
= \tau_0\sum_{n=0}^{N}2^{(h-1)n}
\approx \frac{\tau_0}{1-2^{h-1}},
\end{equation}
where the last relation follows from summation of the geometric progression in
the limit $N\to\infty$.
Thus, in this phenomenological picture, a fluctuation generated at
a small dissipative scale can propagate upscale so that the uncertainty
at the largest eddy grows to the order of its characteristic
velocity amplitude within a few turnover times, independently of how
small the dissipative scale may be.

This observation has a profound consequence: perturbations originating
at arbitrarily small scales can influence the largest scales of the flow
within a finite time horizon. In particular, the amplification and
upscale transport of microscopic noise persist even as the dissipation
scale tends to zero. This mechanism lies at the heart of the physical
picture of spontaneous stochasticity in turbulence.

\section{Solvable toy model: multiscale Arnold's cat}
\label{sec3}

The motivation for introducing a multiscale Arnold's cat model~\cite{mailybaev2023spontaneously} comes directly
from the classical phenomenology of fully developed turbulence.
The purpose of the this model is to illustrate this scenario in the
simplest possible setting.
Rather than attempting to model the full complexity of turbulent dynamics, we
introduce a solvable multiscale toy model that captures three essential
ingredients: (i) a hierarchy of scales with decreasing characteristic times,
(ii) local interactions between neighboring scales, and (iii) chaotic
amplification of perturbations.
The resulting multiscale Arnold's cat model provides a transparent laboratory
in which the emergence of spontaneous stochasticity, universality, and
finite-time propagation across scales can be analyzed explicitly.

\subsection{Ideal model}

Motivated by the phenomenological discussion of the previous section, we now
introduce an idealized multiscale model that captures the essential mechanism of
finite-time propagation across scales.
For simplicity, we take the H\"older exponent to be $h=0$, corresponding to
velocity increments that are independent of scale.
In this case, the hierarchy of spatial scales and the associated turnover times
are given by
\begin{equation}
\ell_n = 2^{-n}, \quad \tau_n = 2^{-n}, \quad n\in\mathbb{Z}_+=\{0,1,2,\dots\}.
\end{equation}
This choice preserves the key feature emphasized earlier: a geometric decrease
of characteristic interaction times across scales.

\begin{figure}[tp]
\centering
\includegraphics[width=0.8\textwidth]{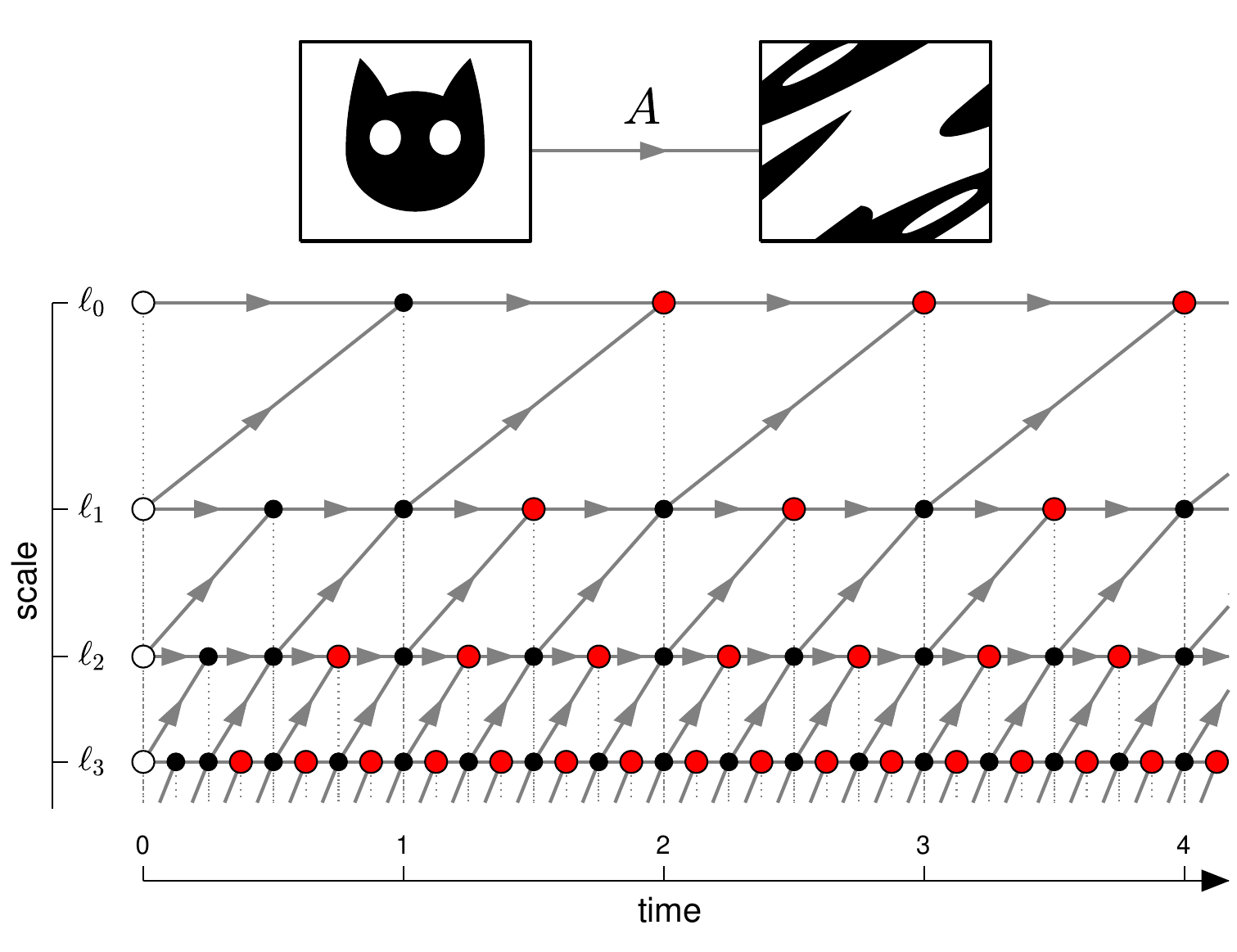}
\caption{Structure of the multi-scale lattice with the variables $u_n(t)$ corresponding to scales $\ell_n$ and discrete times $t \in \tau_n \mathbb{Z}^+$. Gray arrows represent the Arnold's cat map (shown on the top of the figure), which appear in the coupling relation (\ref{eq:cat_multiscale}) and correspond to one turn-over time $\tau_n$. White circles correspond to initial conditions. Red circles denote the variables taking arbitrary values in Theorem~\ref{thm:nonuniqueness}. Small black circles denote the remaining variables, which are uniquely determined by dynamical relations in terms of the white and red ones.}
\label{fig2}
\end{figure}

The resulting multiscale structure can be viewed as a dynamics defined on a
self-similar (monofractal) space--time lattice,
\begin{equation}
\mathcal{L}=\{(n,t): n\in\mathbb{Z}_+,~ t\in \tau_n\mathbb{Z}_+\}.
\label{eq:flatt}
\end{equation}
As illustrated in Fig.~\ref{fig2}, this lattice is invariant under the discrete scaling
transformation
\begin{equation}
t\mapsto 2t, \quad n\mapsto n+1,
\label{eq:scaling_sym_cat}
\end{equation}
which represents a simultaneous rescaling of time and a shift to smaller spatial
scales, $\ell_n\mapsto \ell_{n+1}$. This scale invariance, however, is broken at the largest scale $n = 0$.

At each lattice point $(n,t)\in\mathcal{L}$ we introduce a state variable
$u_n(t)$ taking values on the two-dimensional torus
$\mathbb{T}^2=\mathbb{R}^2/\mathbb{Z}^2$.
The variable $u_n(t)$ may be interpreted as a phase associated with an eddy of
characteristic size $\ell_n$ at time $t$.
The multiscale dynamics is defined by the deterministic update rule
\begin{equation}
u_n(t+\tau_n) = A \big(u_n(t) + u_{n+1}(t) \big),
\label{eq:cat_multiscale}
\end{equation}
where $A: \mathbb{T}^2 \mapsto \mathbb{T}^2$ denotes the Arnold cat map
\begin{equation}
A:\ (x,y)\mapsto (2x+y,\ x+y)\ \mathrm{mod}\ 1.
\label{eq:cat_matrix}
\end{equation}
This map is linear, hyperbolic, invertible and area-preserving.
Equation~\eqref{eq:cat_multiscale} couples each scale $n$ to itself and to the
immediately smaller scale $n+1$, while preserving the discrete scaling symmetry
\eqref{eq:scaling_sym_cat}; see Fig.~\ref{fig2}.

We prescribe deterministic initial conditions at all scales,
\begin{equation}
u_n(0)=a_n,\quad n\in\mathbb{Z}_+.
\label{eq:cat_IC}
\end{equation}
Despite being formally deterministic, the infinite system
\eqref{eq:cat_multiscale}--\eqref{eq:cat_IC} exhibits a fundamental lack of
well-posedness.
Information propagates from arbitrarily small scales toward larger ones, and the
dynamics does not uniquely determine the evolution from the given initial data.
This lack of uniqueness can be stated precisely as follows.

\begin{theorem}[Non-uniqueness of solutions {\cite{mailybaev2023spontaneously}}]
\label{thm:nonuniqueness}
For any prescribed initial condition \eqref{eq:cat_IC}, the infinite multiscale
system \eqref{eq:cat_multiscale} admits uncountably many distinct solutions
defined for all times $t\ge 0$.
Specifically, one may choose arbitrary values of $u_n(t)$ at times $t=m\tau_n$, where
$m\ge 2$ for $n=0$ and $m\ge 3$ odd for $n\ge 1$; see the red nodes in Fig.~\ref{fig2}.
Once these values are prescribed, the dynamical relations
\eqref{eq:cat_multiscale} uniquely determine $u_n(t)$ at all remaining
(black) nodes of the lattice.
\end{theorem}

This intrinsic non-uniqueness reflects an ``information deficit'' flowing from
small to large scales and provides a clean analogue of the non-uniqueness of weak
solutions encountered in ideal fluid models.
As we shall see, this feature plays a central role in the emergence of
spontaneous stochasticity when suitable regularizations are introduced.

\subsection{Regularization, ideal limit, and spontaneous stochasticity}

As in the case of ideal fluid equations, the lack of well-posedness described in
Theorem~\ref{thm:nonuniqueness} suggests that physically relevant solutions
should be obtained as limits of suitably regularized models.
In the present setting, such a regularization can be introduced in a natural
and transparent manner.

\begin{figure}[tp]
\centering
\includegraphics[width=0.85\textwidth]{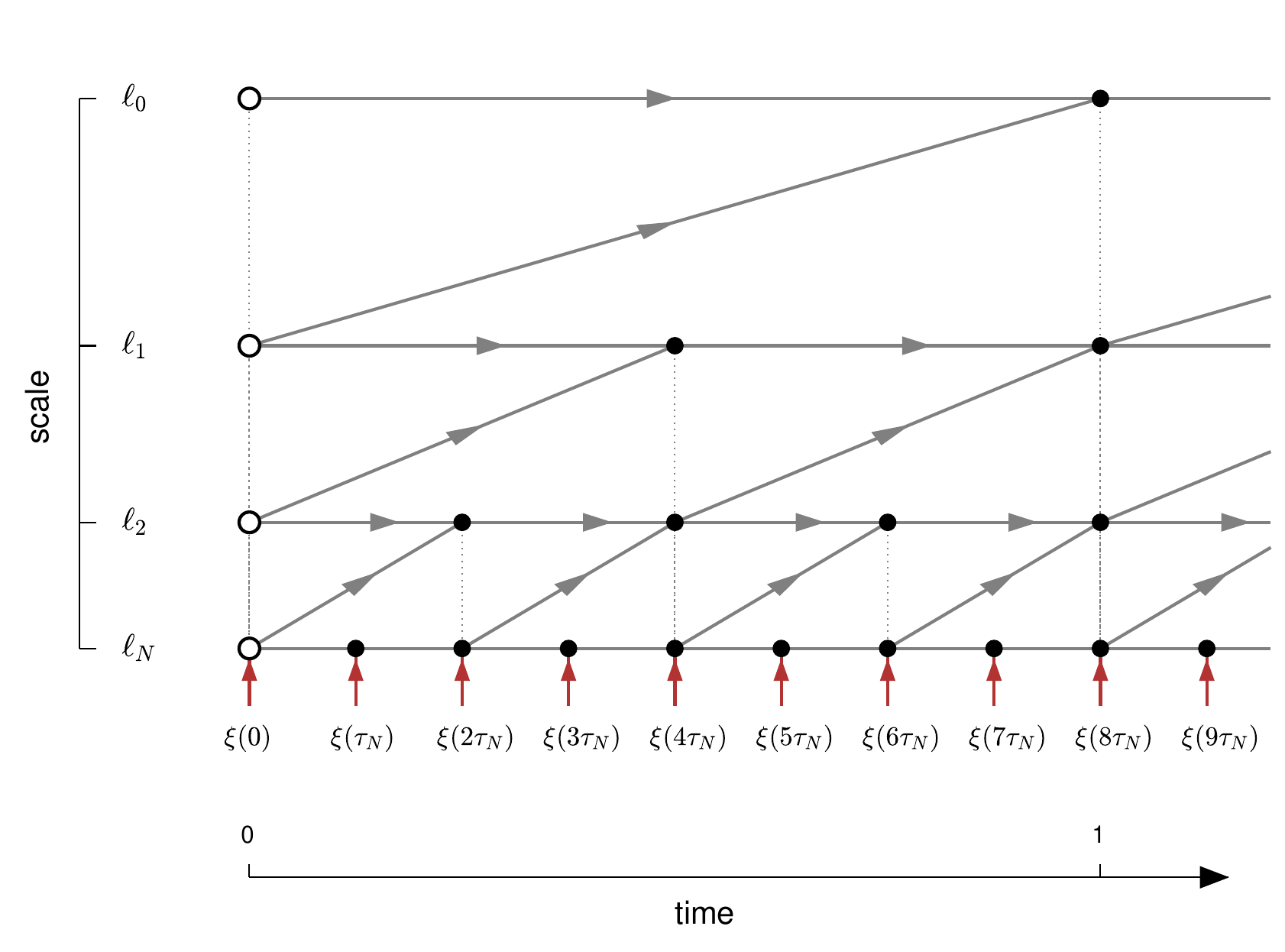}
\caption{Multiscale system truncated at the cutoff scale $\ell_N$.
Only the variables $u_n(t)$ for $n=0,\ldots,N$ are retained,
while $u_n(t)\equiv 0$ for $n>N$.
At the smallest active scale $\ell_N$, random forcing $\xi(t)$
is injected at discrete times $t=m\tau_N$ (red arrows),
modifying the evolution according to Eq.~\eqref{eq:cat_multiscale_cutoff}.}
\label{figB}
\end{figure}

We introduce a cutoff at scale $N$ by truncating the multiscale system at
$n=N$, retaining only the variables $u_n(t)$ for $n=0,\ldots,N$ and setting
$u_n(t)\equiv 0$ for $n>N$. 
Here and in what follows, the notation $0$ denotes the origin $(0,0)$ of the torus $\mathbb{T}^2$.
At the cutoff scale, the evolution equation is modified to
\begin{equation}
u_N(t+\tau_N) = A \big( u_N(t) + \xi(t) \big),
\label{eq:cat_multiscale_cutoff}
\end{equation}
where $\{\xi(t)\}$ are independent and identically distributed random variables
defined at times $t=m\tau_N$; see Fig.~\ref{figB}.

This cutoff regularization naturally mimics the role of viscosity in turbulence,
which damps the dynamics below a viscous scale, and
the random variables $\xi(t)$ provide a source of microscopic noise.
The \emph{ideal limit} is then formulated as $N\to\infty$, in which both the
cutoff and the noise are shifted to arbitrarily small scales and formally
disappear.
In this limit, the regularized dynamics reduces to the ideal system introduced
in the previous subsection.

For any fixed cutoff $N$, the resulting system is finite-dimensional and
well-posed.
In particular, for each initial condition \eqref{eq:cat_IC}, the truncated
dynamics defines a unique probabilistic solution, i.e.\ a stochastic process
defined for all times $t\ge 0$. 
More precisely, the variable $u_n(t)$ at each node can be written as a
linear combination of the initial data $a_m$ for $m=n,\ldots,N$ and of
the noise variables $\xi(s)$ for $s<t$; see Fig.~\ref{figB}.
The central result is that, as the cutoff is removed, these stochastic
solutions converge to a universal limiting stochastic process.

\begin{theorem}[Spontaneously stochastic limit
{\cite{mailybaev2023spontaneously}}]
\label{thm:convergence}
Fix an arbitrary initial state~\eqref{eq:cat_IC}.
Let $\big\{u_n^{(N)}(t)\big\}_{(n,t)\in\mathcal{L}}$ denote the stochastic solution of the system with cutoff at scale $N$.
As $N\to\infty$, these solutions converge in law with respect to the
product topology on the configuration space.
The limiting distribution is independent of the specific choice of the
noise $\xi(t)$, provided that its law is absolutely continuous. 
Moreover, the limit coincides with the class of nonunique solutions described in
Theorem~\ref{thm:nonuniqueness}, in which the free variables $u_n(t)$ are
independent and uniformly distributed on $\mathbb{T}^2$.
\end{theorem}

We remark that Theorem~\ref{thm:convergence} was proved in
\cite{mailybaev2023spontaneously} for the case where the noise acts only at the
initial time.
However, owing to the linearity of the dynamics, the argument extends directly
to the temporally distributed noise considered here.

It follows that the limiting process described in
Theorem~\ref{thm:convergence} satisfies all four defining properties of
spontaneous stochasticity.

\emph{(i) Convergence:}
The probability law of the regularized dynamics converges as
the cutoff scale is removed.

\emph{(ii) Stochasticity:}
The limiting law is intrinsically non-deterministic.
Moreover, it is ``maximally'' random in the sense that all 
non-unique solutions from Theorem~\ref{thm:nonuniqueness} are selected with equal probability.

\emph{(iii) Spontaneity:}
The limiting stochastic process is supported on nonunique solutions of the
deterministic ideal system with deterministic initial conditions.

\emph{(iv) Universality:}
Different absolutely continuous noise distributions imposed at
the cutoff scale lead to the same ideal-limit law.

We conclude that the stochastic regularization described above induces a selection
principle for the non-unique ideal dynamics.
In the ideal limit, a universal probability distribution emerges, assigning
well-defined statistics to the degrees of freedom associated with
arbitrarily small scales.
This randomness corresponds precisely to the uncountable family
of deterministic solutions admitted by the ideal system.
In this sense, spontaneous stochasticity provides a probabilistic resolution of
non-uniqueness, analogous to the role played by entropy conditions in
conservation laws and by admissibility criteria in weak formulations of fluid
equations.

\section{RG framework for the multiscale Arnold's cat}
\label{sec4}

We now reinterpret this selection mechanism within a
renormalization-group (RG) framework~\cite{mailybaev2023spontaneous,mailybaev2025rg},
where the ideal-limit probability distribution emerges as a fixed point of a suitable RG
transformation acting on multiscale transition kernels.

\subsection{RG relation in the absence of noise}

Let us consider a regularized system with cutoff $N$.
It is instructive to first analyze the deterministic case, in which no noise is
present, i.e.\ $\xi(t)\equiv 0$. At integer times $t \in \mathbb{Z}_+$, we denote the full state of the system by
\begin{equation}
u(t) = \big(u_0(t),\,u_1(t),\,u_2(t),\ldots \big) \in (\mathbb{T}^2)^{\mathbb{Z}_+}.
\label{eq:fullstate}
\end{equation}
Similarly, we denote the full sequence of initial conditions as $a = (a_0,a_1,a_2,\ldots)$.
Let $\varphi^{(N)}:(\mathbb{T}^2)^{\mathbb{Z}_+} \mapsto (\mathbb{T}^2)^{\mathbb{Z}_+}$ be the deterministic map defining the unit-time
evolution of the system with the cutoff $N$ as
\begin{equation}
\varphi^{(N)}(a) = u(1).
\label{eq:phi_det}
\end{equation}
We call it the flow map.
The evolution to any integer time $t$ is then given by the
$t$-fold composition of $\varphi^{(N)}$. For $N = 0$, the dynamics is limited to the scale $n = 0$ 
and has the form (\ref{eq:cat_multiscale_cutoff}) with zero noise term. 
Hence, the respective flow map is simply
\begin{equation}
\varphi^{(0)}(a) = (Aa_0,0,0,\ldots).
\label{eq:phi0}
\end{equation}

\begin{figure}[t]
\centering
\includegraphics[width=0.85\textwidth]{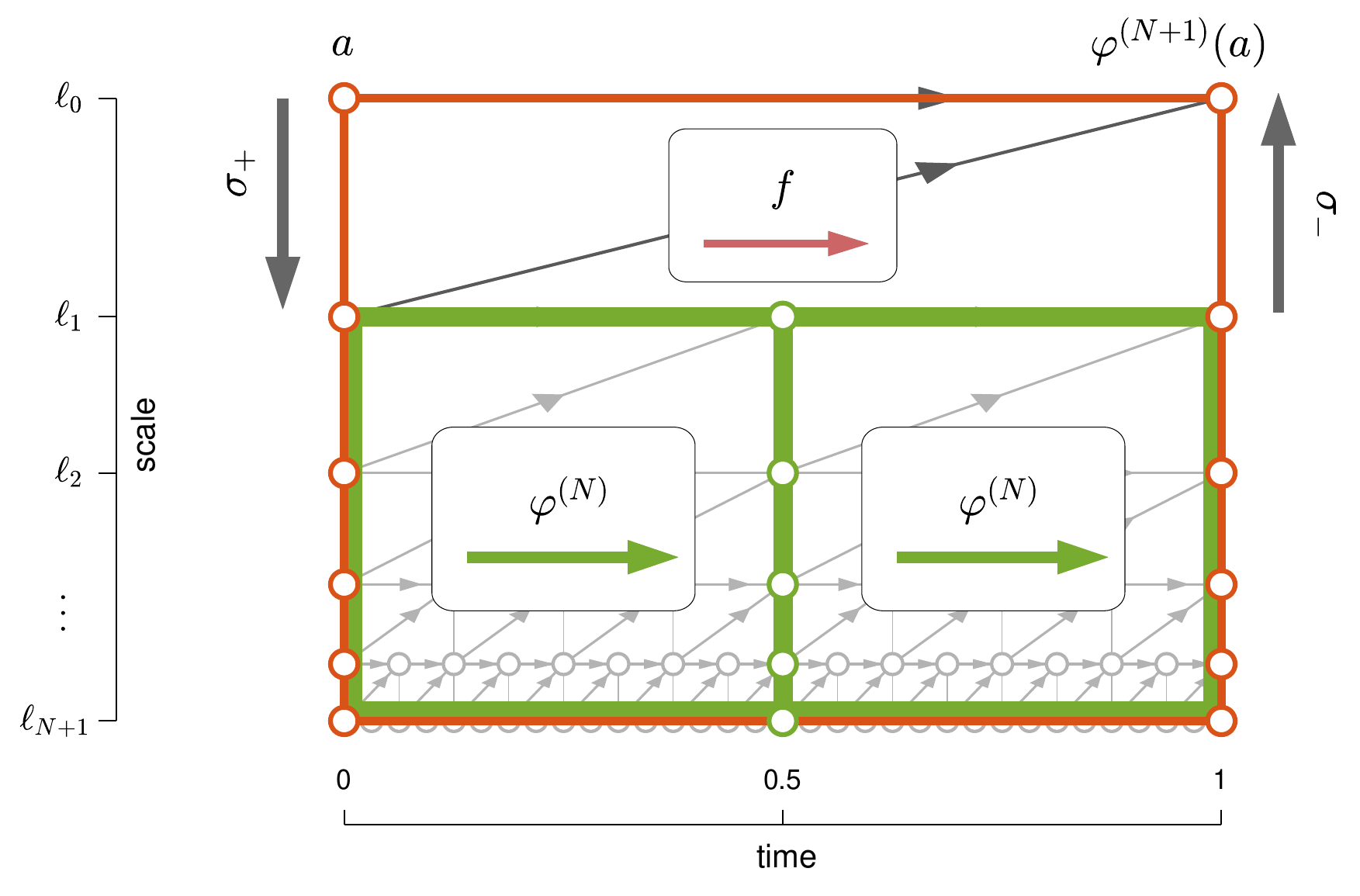}
\caption{Block decomposition underlying the RG relation~\eqref{eq:RGdet}.
The unit-time evolution of the cutoff-$(N+1)$ system is represented as two
consecutive unit-time evolutions $\varphi^{(N)}$ of the cutoff-$N$ system, followed by an update
of the largest scale via the map $f$.
Scale alignment is achieved using the shift maps $\sigma_+$ and $\sigma_-$.}
\label{fig3}
\end{figure}

For further analysis, we introduce the shift maps
\begin{equation}
\sigma_+(u) = (u_1,u_2,\ldots), \quad
\sigma_-(u) = (0,u_0,u_1,\ldots),
\label{eq:maps}
\end{equation}
where $\sigma_+$ represents a shift toward larger scales and
$\sigma_-$ a shift toward smaller scales with zero padding.
We also define the large-scale update map
\begin{equation}
f(u) = \big(A (u_0 + u_1), 0,0,\ldots\big),
\label{eq:ACmap}
\end{equation}
which governs the evolution of the largest scale over one unit time step.
Then the RG relation, which expresses the flow map $\varphi^{(N+1)}$ of the
cutoff-$(N+1)$ system explicitly in terms of $\varphi^{(N)}$,
can be written as~\cite{mailybaev2023spontaneous}
\begin{equation}
\varphi^{(N+1)}
=
\sigma_-\circ \varphi^{(N)} \circ \varphi^{(N)} \circ \sigma_+  + f.
\label{eq:RGdet}
\end{equation}
This relation follows from a decomposition of the dynamics into three elementary
parts, schematically illustrated in Fig.~\ref{fig3}.
Here the evolution over unit-time step of the cutoff-$(N+1)$ system consists of
two consecutive blocks, each equivalent to a unit-time evolution of the
cutoff-$N$ system, combined with the update of the largest scale governed by
the map $f$.
The shift maps $\sigma_+$ and $\sigma_-$ are used to align scales and to restore
the original lattice structure.

Equation~\eqref{eq:RGdet} thus defines a deterministic RG transformation acting
on the space of flow maps, and starting from the initial state (\ref{eq:phi0}). 
However, these flow maps fail to converge as $N\to\infty$
\cite{mailybaev2023spontaneous}.
As we show in the next subsection, this deterministic RG relation admits a
natural extension to the stochastic setting, in which it acts on Markov kernels
rather than on maps and leads to a fixed-point attractor corresponding to
spontaneous stochasticity.

\subsection{RG relation for Markov kernels}

Consider now a regularized system with cutoff $N$ in the presence of noise $\xi(t)$. Recall that our noise 
is statistically independent and identically distributed in times $t = m\tau_N$, $m \in \mathbb{Z}_+$.
Hence, the stochastic dynamics $u(t)$ observed at integer times $t = \mathbb{Z}_+$ represents a
Markov chain.
The one-step transition probabilities are described by the Markov kernel
\begin{equation}
\Phi^{(N)}(\mathcal{U}\,|\,a)
:= \mathbb{P}\!\left(u(1)\in\mathcal{U}\,\big|\,u(0)=a\right),
\label{eq:flowkernel_def}
\end{equation}
which gives the probability that the state starting from $u(0)=a$ belongs to a measurable
set $\mathcal{U} \subset (\mathbb{T}^2)^{\mathbb{Z}_+}$ at time $t=1$.
The same kernel governs the transition from $u(t)$ to $u(t+1)$ for any integer time.
As a technical remark, we note that the state space
$u\in(\mathbb{T}^2)^{\mathbb{Z}_+}$ is infinite-dimensional, and that a rigorous
probabilistic formulation relies on the infinite-product topology; see
\cite{mailybaev2023spontaneous}.

In the case $N=0$, the stochastic dynamics is confined to the single scale
$n=0$ and is governed by the update rule~\eqref{eq:cat_multiscale_cutoff}.
In this case, the one-step transition kernel can be written explicitly as a
product measure,
\begin{equation}
\Phi^{(0)}(\cdot\,|\,a)
=
\nu_{a_0}\times \delta_0 \times \delta_0 \times \cdots,
\label{eq:phi0st}
\end{equation}
where $\delta_0$ denotes the Dirac measure at the origin for the components
$u_n(1)$ with $n\ge1$.
The probability measure $\nu_{a_0}$ depends only on the initial component $a_0$ and
is given as the pushforward
\begin{equation}
\nu_{a_0} = (g_{a_0})_\sharp \mu_\xi, \quad
g_{a_0}(\xi) = A (a_0+\xi),
\label{eq:phi0stn0}
\end{equation}
where $\mu_\xi$ denotes the probability distribution of the noise variable
$\xi(0)$.

The RG relation provides an explicit expression for the kernel
$\Phi^{(N+1)}$ in terms of $\Phi^{(N)}$.
This relation can be deduced directly from the deterministic RG relation
\eqref{eq:RGdet} by replacing each map with the corresponding Markov kernel.
Under this correspondence, the composition of maps becomes the composition of
kernels, while the sum of maps is replaced by the convolution operation,
denoted by the asterisk~\cite{mailybaev2023spontaneous}.
For the deterministic maps $\sigma_\pm$ and $f$, one introduces the associated
Dirac Markov kernels and denote them by the corresponding capital letters.
For example, the kernel $\Sigma_+(\mathcal{U}\,|\,a)$ equals $1$ if
$\sigma_+(a)\in\mathcal{U}$ and $0$ otherwise.
The same convention applies to $\Sigma_-$ and $F$.

We can now state the stochastic RG relation for the multiscale Arnold's cat model.

\begin{theorem}[Stochastic RG relation {\cite{mailybaev2023spontaneous}}]
\label{thm:RG}
For the multiscale Arnold's cat model, the Markov kernels at successive cutoffs
satisfy
\begin{equation}
\Phi^{(N+1)} =
\big(
\Sigma_- \circ \Phi^{(N)} \circ \Phi^{(N)} \circ \Sigma_+
\big) * F.
\label{eq:RGAC}
\end{equation}
\end{theorem}

Recall that the composition of kernels is defined by
\begin{equation}
\Phi_2\circ\Phi_1(\mathcal{U}\,|\,a)
=
\int \Phi_2(\mathcal{U}\,|\,v)\,\Phi_1(dv\,|\,a),
\label{eq:kernel_comp_def}
\end{equation}
where the integration is taken over the intermediate state $v$. 
The convolution with the Dirac kernel $F$ is given by
\begin{equation}
\Phi * F\,(\mathcal{U}\,|\,a)
= \Phi(\mathcal{U}_{f(a)} |\,a), \quad
\mathcal{U}_{f(a)} = \{ u: u+f(a) \in \mathcal{U} \}.
\label{eq:kernel_conv_def}
\end{equation}

It is convenient to rewrite the RG relation in operator form as
\begin{equation}
\Phi^{(N+1)} = \mathcal{R}[\Phi^{(N)}],
\label{eq:RGrelation}
\end{equation}
where the RG operator $\mathcal{R}$ acts on Markov kernels $\Phi$ according to
\begin{equation}
\mathcal{R}[\Phi]
=
\big( \Sigma_- \circ \Phi \circ \Phi \circ \Sigma_+ \big) * F.
\label{eq:RGoperator}
\end{equation}
Thus, the change of the cutoff scale is represented as an RG dynamics in the
space of Markov kernels.
Within this framework, the inviscid (ideal) limit corresponds to an attractor of
the RG dynamics, that is, to a fixed point of the operator $\mathcal{R}$.
Note that the RG dynamics is initialized by the kernel~\eqref{eq:phi0st}, which
depends on the specific distribution of the microscopic noise.
In contrast, the RG operator $\mathcal{R}$ itself depends only on the coupling
structure of the ideal system and is independent of the noise statistics.

\subsection{RG fixed point and its stability}

The ideal limit $N\to\infty$ corresponds, in RG language, to the infinite
iteration of the map $\mathcal{R}$.
A kernel $\Phi^\ast$ satisfying
\begin{equation}
\Phi^* = \mathcal{R}[\Phi^*]
\label{eq:RG_fixed_point}
\end{equation}
is an RG fixed point.
Such a fixed point represents a scale-invariant stochastic dynamics on the
multiscale lattice. 
According to Theorem~\ref{thm:convergence} we know this fixed point explicitly: 
the state $u(1)$ has the deterministic first component $u_0(1) = A (a_0+a_1)$, while the remaining components $u_n(1)$ for $n \ge 1$ are uniformly and independently distributed~\cite{mailybaev2023spontaneously}. This can be written for a given initial state $a$ as
\begin{equation}
\Phi^*(\cdot\,|\,a) = \delta_{A (a_0+a_1)} \times \lambda \times \lambda \times \cdots,
\label{eq:RG_fixed_point_form}
\end{equation}
which represents a product of the Dirac measure for the first component and uniform (Lebesgue) measures $\lambda$ for the remaining components. Using expression (\ref{eq:RGoperator}), it is straightforward to check that $\Phi^*$ is indeed the fixed point of the RG operator. 

The convergence result of Theorem~\ref{thm:convergence} implies that the sequence
$\{\Phi^{(N)}\}_{N \ge 0}$ converges, as $N\to\infty$, to the fixed point $\Phi^*$
for any initial kernel~\eqref{eq:phi0st} generated by an absolutely continuous
noise term. This strongly suggests that $\Phi^*$ is an attractor of the
RG dynamics. A natural way to investigate this stability property is to
linearize the RG operator about the fixed point.
We write
\begin{equation}
\Phi^{(N)} = \Phi^* + \delta\Phi^{(N)},
\label{eq:RG_perturbation}
\end{equation}
where $\delta\Phi^{(N)}$ is a signed kernel describing a deviation from the attractor.
Assuming that these deviations are small for sufficiently large $N$, the
linearized RG relation~\eqref{eq:RGrelation} takes the form
\begin{equation}
\delta\Phi^{(N+1)} = D\mathcal{R}[\delta\Phi^{(N)}],
\label{eq:Lin_RGrelation}
\end{equation}
where $D\mathcal{R}$ denotes the Fr\'echet derivative of the RG operator.
From~\eqref{eq:RGoperator} one obtains
\begin{equation}
D\mathcal{R}[\delta\Phi]
=
\big( \Sigma_- \circ \Phi^* \circ \delta\Phi \circ \Sigma_+ \big) * F
+
\big( \Sigma_- \circ \delta\Phi \circ \Phi^* \circ \Sigma_+ \big) * F.
\label{eq:Lin_RGoperator}
\end{equation}

Seeking solutions of the form
$\delta\Phi^{(N)} = \rho^N \Psi$, the linearized RG relation~\eqref{eq:Lin_RGrelation}
reduces to the eigenvalue problem
\begin{equation}
D\mathcal{R}[\Psi] = \rho\, \Psi.
\label{eq:Lin_RGeig}
\end{equation}
The eigenvalues $\rho$ determine the stability properties of the RG
fixed point, separating unstable modes ($|\rho|>1$) from stable ones
($|\rho|<1$).

\begin{figure}[tp]
\centering
\includegraphics[width=0.75\textwidth]{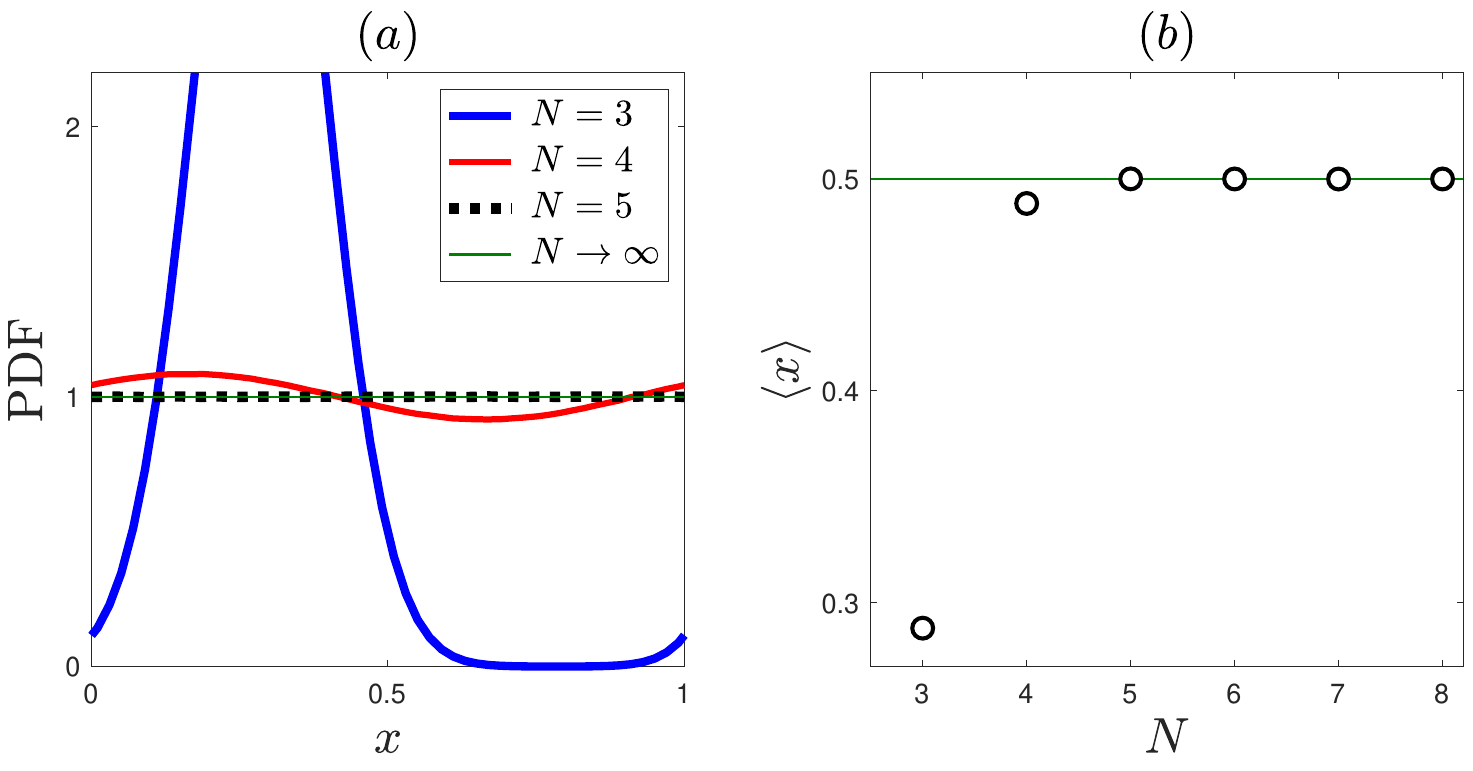}
\caption{(a) Probability density functions and (b) mean values for the first
component of the random variable $u_1(1) = (x,y)$.
The thin green line corresponds to the uniform distribution associated with the
ideal limit.
The data are based on $10^8$ samples from numerical simulations with initial
conditions $a_n = (2^{n/2},\,2^{n/3})$ and normally distributed noise
$\xi(t)$ with zero mean and variance $\sigma^2 = 10^{-6}$.}
\label{figAC}
\end{figure}

At present, no analytical solution of the eigenvalue problem~\eqref{eq:Lin_RGeig}
is available. However, numerical simulations (see Fig.~\ref{figAC}) suggest
that the RG fixed point may be superattracting, in the sense that the
convergence of $\Phi^{(N)}$ to $\Phi^*$ appears to be superexponential.
At the linear level, this would imply that all eigenvalues satisfy $\rho=0$.
If confirmed, such superattracting behavior should be regarded as an artifact
of the simplified linear interaction structure of the present model.
More sophisticated (and nonlinear) models defined on the same fractal lattice
(see~\cite{mailybaev2025rg}) exhibit nonzero eigenvalues
$0<|\rho|<1$, corresponding to stable directions of the RG flow.

\section{Feigenbaum RG equation on the fractal lattice}
\label{sec5}

We now place the RG formulation developed above in a broader context by relating
it to classical renormalization-group constructions.
Our aim is to show that several well-known RG theories can be recovered from
dynamics defined on the same self-similar (fractal) space--time lattice.
The distinction between these theories lies in the local dynamical rules imposed
on the lattice, which define the ideal system and, consequently, the associated
RG dynamics.

We begin with the Feigenbaum--Cvitanovi\'c functional equation~\cite{feigenbaum1978quantitative},
introduced independently by Coullet and Tresser~\cite{coullet1978iterations}.
Their theory describes universal scaling properties in one-dimensional
discrete-time dynamics
\begin{equation}
x_{m+1} = f(x_m), \quad m=0,1,2,\ldots,
\label{eq:Feig_map}
\end{equation}
where $x_m\in\mathbb{R}$ and $f:\mathbb{R}\to\mathbb{R}$ is a nonlinear map.
Near the onset of chaos, the dynamics undergoes a period-doubling accumulation,
characterized by self-similarity under the combined action of time iteration
and rescaling of phase space.
Rather than following the standard derivation of the Feigenbaum equation, we
show here how the same scaling relation emerges naturally from dynamics defined
on the fractal lattice.

\subsection{Representation on the fractal lattice}

We consider the same fractal lattice $\mathcal{L}$ from Eq.~(\ref{eq:flatt}) and introduce a real-valued
field $u_n(t)\in\mathbb{R}$ defined on its nodes.
The discrete-time dynamics~\eqref{eq:Feig_map} is assigned to the smallest
resolved scale $n=N$, for which a single iteration corresponds to the turnover
time $\tau_N=2^{-N}$.
Specifically, we define
\begin{equation}
u_N(t+\tau_N) = f\big(u_N(t)\big).
\label{eq:Feig_cutoff}
\end{equation}
The correspondence with Eq.~\eqref{eq:Feig_map} is given by
$u_N(m\tau_N)=x_m$.
As before, we set $u_n(t)\equiv0$ for all $n>N$.

\begin{figure}[tp]
\centering
\includegraphics[width=0.9\textwidth]{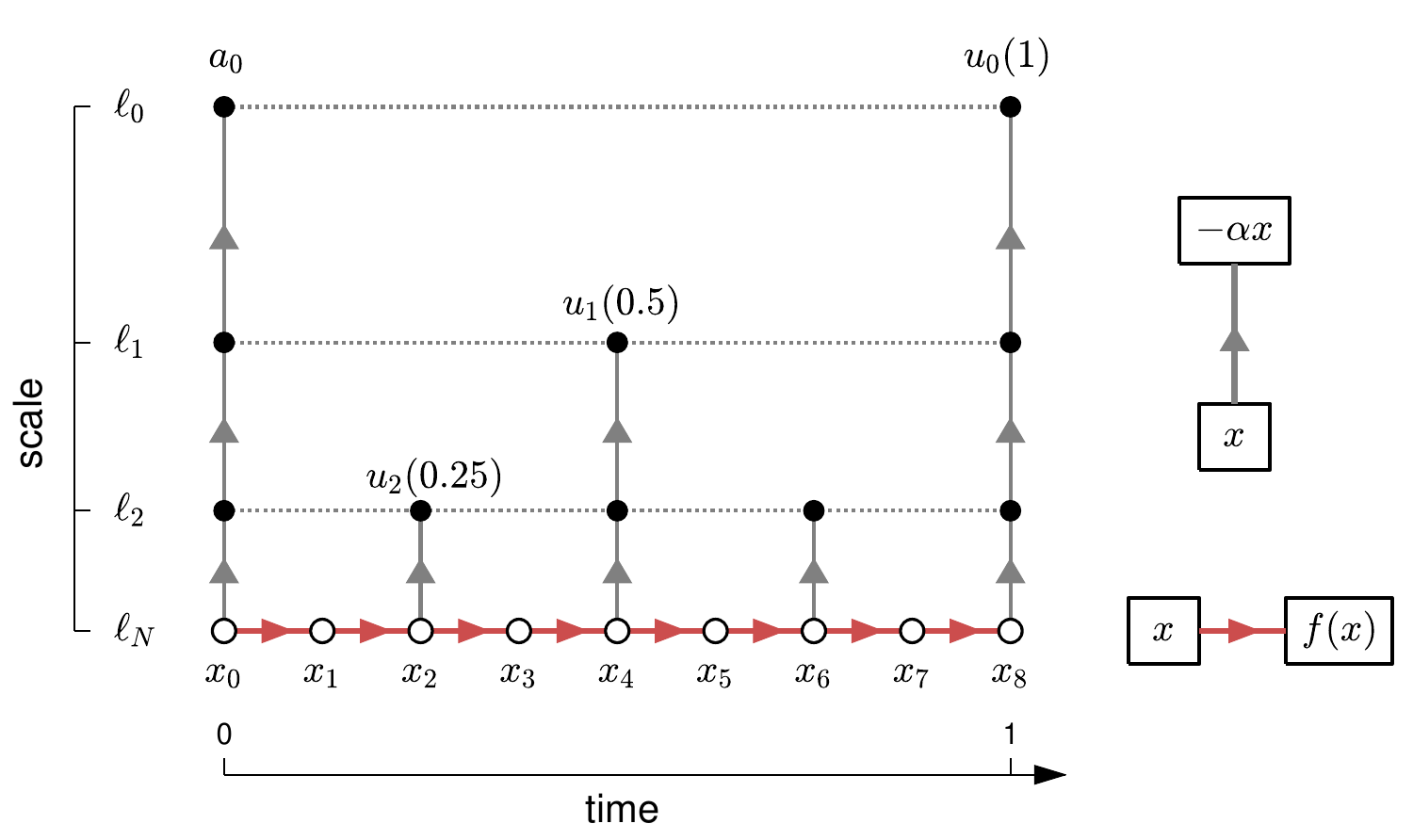}
\caption{Dynamics on the fractal lattice for the Feigenbaum period-doubling relation.
The original dynamical sequence $\{x_m\}_{m\ge0}$ is assigned to the cutoff scale
$N$.
Vertical arrows denote the scaling relations~\eqref{eq:Feig_scaling} between the
variables $u_n(t)$ at different scales.}
\label{fig4}
\end{figure}

At larger scales $n<N$, the dynamics is purely self-similar and consists
only of rescaling between neighboring nodes.
As illustrated schematically in Fig.~\ref{fig4}, we impose the relation
\begin{equation}
u_n(t) = -\alpha\,u_{n+1}(t),
\label{eq:Feig_scaling}
\end{equation}
where $\alpha>0$ is a fixed parameter, known as the Feigenbaum scaling constant.
Applying~\eqref{eq:Feig_scaling} to the initial states $u_n(0)=a_n$ yields
$a_n=(-\alpha)^{-n}a_0$, so that the initial condition is completely determined
by a single value,
\begin{equation}
u_0(0)=a_0.
\label{eq:Feig_IC}
\end{equation}
No noise is introduced in this formulation.

The scaling rule~\eqref{eq:Feig_scaling} plays the role of a deterministic coupling
between scales in the \emph{ideal} system.
Together with the cutoff-scale dynamics~\eqref{eq:Feig_cutoff} and the initial
condition~\eqref{eq:Feig_IC}, it completely defines the evolution on the
lattice.
As a result, the entire field $u_n(t)$ is uniquely determined by the initial
value $a_0$ and the map $f$.

Let us restrict attention to integer times $t=m\in\mathbb{Z}_+$.
The scaling relation~\eqref{eq:Feig_scaling} implies that all variables
$u_n(m)=(-\alpha)^{-n}u_0(m)$, for $n=1,\ldots,N$, are uniquely expressed in terms
of the largest-scale variable $u_0(m)$.
Consequently, the dynamics reduces to an effective evolution at the largest
scale $n=0$.
This evolution is governed by the flow map corresponding to a unit-time step,
which we introduce next.

\subsection{RG dynamics and its fixed point}

For a given cutoff $N$, we define a flow map
$\varphi^{(N)}:\mathbb{R}\to\mathbb{R}$ by
\begin{equation}
\varphi^{(N)}(a_0) = u_0(1),
\label{eq:Feig_phi}
\end{equation}
that is, $\varphi^{(N)}$ maps the initial condition at the largest scale to the
value of the largest-scale variable after one large-scale turnover time
$\tau_0=1$; see Fig.~\ref{fig4}.
For $N=0$, it follows directly from~\eqref{eq:Feig_cutoff} that
\begin{equation}
\varphi^{(0)} = f.
\label{eq:Feig_phi0}
\end{equation}
For $N>0$, the flow map $\varphi^{(N)}$ represents 
\begin{equation}
\varphi^{(N)}(x)
=
(-\alpha)^N\, f^{\circ 2^N}\!\left(\frac{x}{(-\alpha)^N}\right),
\label{eq:Feig_phi_iter}
\end{equation}
where $f^{\circ 2^N}$ denotes the $2^N$-times composition of $f$ with itself.

\begin{figure}[tp]
\centering
\includegraphics[width=0.6\textwidth]{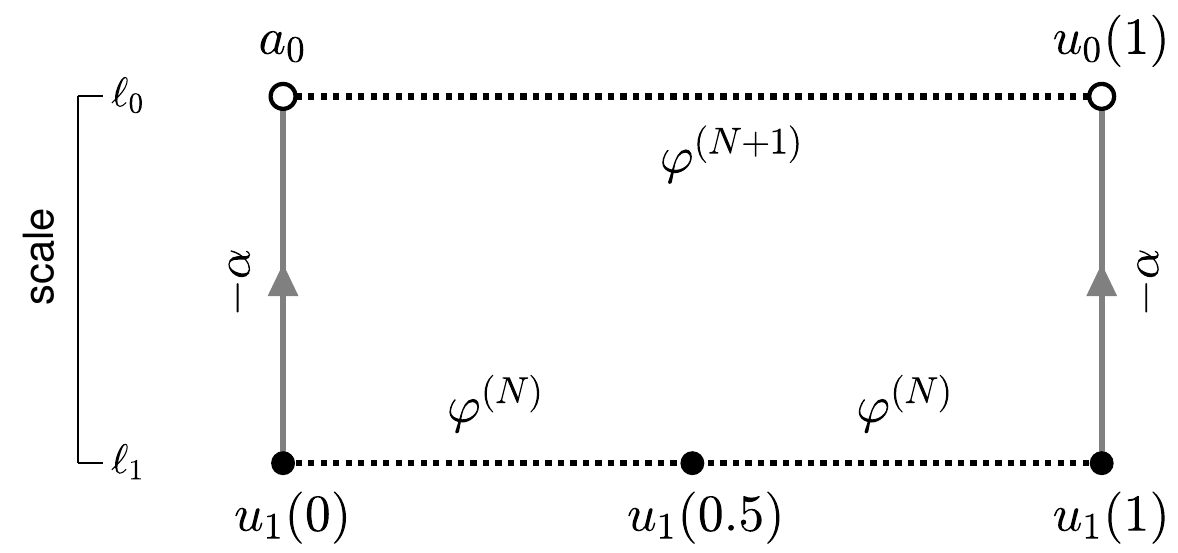}
\caption{Block structure of the cutoff-$(N+1)$ dynamics leading to the Feigenbaum RG
equation (\ref{eq:Feig_RG}) for the flow maps.}
\label{fig5}
\end{figure}

Using the same block decomposition of the lattice as in the multiscale Arnold's
cat model, now adapted to the present setting (see Fig.~\ref{fig5}), one can
relate the maps $\varphi^{(N+1)}$ and $\varphi^{(N)}$.
The scaling relation~\eqref{eq:Feig_scaling} connects the states at times
$t=0$ and $t=1$.
By self-similarity, the turnover-time dynamics at scale $n=1$ of the
cutoff-$(N+1)$ system is governed by the map $\varphi^{(N)}$.
As a result, one obtains
\begin{equation}
\varphi^{(N+1)}(a_0)
=
-\alpha\,\varphi^{(N)}\!\circ\!\varphi^{(N)}(-a_0/\alpha).
\label{eq:Feig_RG}
\end{equation}
This relation can be written compactly as an RG iteration,
\begin{equation}
\varphi^{(N+1)} = \mathcal{R}[\varphi^{(N)}],
\label{eq:Feig_RG_iter}
\end{equation}
where the RG operator $\mathcal{R}$ acts on functions
$\varphi:\mathbb{R}\to\mathbb{R}$ according to
\begin{equation}
\mathcal{R}[\varphi](x)
=
-\alpha\,\varphi\!\circ\!\varphi(-x/\alpha).
\label{eq:Feig_RG_op}
\end{equation}
Note the remarkable structural similarity between this RG operator and the one
arising in the multiscale Arnold's cat model; see
Eqs.~\eqref{eq:RGdet} and~\eqref{eq:RGAC}, where the shift operators
$\sigma_+$ and $\sigma_-$ play a role analogous to scaling transformations in
the state space.

As in the multiscale Arnold's cat model, the ideal limit $N\to\infty$ is thus
reduced to an RG flow in the space of flow maps, starting from the initial
condition~\eqref{eq:Feig_phi0}.
Fixed points of this RG flow satisfy
\begin{equation}
\varphi^\ast = \mathcal{R}[\varphi^\ast],
\label{eq:Feig_fixed_point}
\end{equation}
which is precisely the Feigenbaum--Cvitanovi\'c functional equation.
Its nontrivial solution $\varphi^\ast$ exists for a specific value of the
Feigenbaum scaling constant, $\alpha \approx 2.5029$, and is known as the
Feigenbaum universal function.

\subsection{Stability and universality}

To analyze the stability of this fixed point, one linearizes the RG operator
about $\varphi^\ast$ by writing $\varphi^{(N)}=\varphi^\ast+\delta\varphi^{(N)}$.
This yields the linearized RG iteration
\begin{equation}
\delta\varphi^{(N+1)} = D\mathcal{R}[\delta\varphi^{(N)}],
\label{eq:Feig_fixed_point_lin}
\end{equation}
where the Fr\'echet derivative of the RG operator is given by
\begin{equation}
D\mathcal{R}[\psi](x)
=
-\alpha\,\psi\big(\varphi^\ast(-x/\alpha)\big)
-\alpha\,\varphi^{\ast\prime}\big(\varphi^\ast(-x/\alpha)\big)\,\psi(-x/\alpha),
\end{equation}
and the prime denotes differentiation with respect to the argument.
Seeking solutions of the form $\delta\varphi^{(N)}=\rho^N\psi$ leads to the
eigenvalue problem
\begin{equation}
D\mathcal{R}[\psi] = \rho\,\psi.
\label{eq:Feig_fixed_point_eig}
\end{equation}
The eigenvalues $\rho$ determine the stability properties of the Feigenbaum RG
fixed point, separating unstable modes ($|\rho|>1$) from stable ones
($|\rho|<1$).

In contrast to the multiscale Arnold's cat model, the Feigenbaum fixed point is
not stable.
After disregarding unstable directions that can be eliminated by appropriate
normalization conditions, the RG spectrum contains a single genuinely unstable
(so-called \emph{relevant}) direction, with eigenvalue
$\rho=\delta\approx4.6692$.
Together with the scaling constant $\alpha$, this eigenvalue determines the
universal scaling properties of the period-doubling
cascade~\cite{feigenbaum1983universal}.

We observe that both the fixed point and its stability properties are
completely determined by the RG operator~\eqref{eq:Feig_RG_op}, namely,
by the deterministic dynamical rules prescribed on the fractal lattice.
From this perspective, the transition from the multiscale Arnold's cat model
to the Feigenbaum universality class amounts simply to a modification of the
local dynamical rules (together with a change of the phase space from
$\mathbb{T}^2$ to $\mathbb{R}$), while the underlying self-similar lattice
structure remains unchanged.

\section{RG formulation of the central limit theorem}
\label{sec6}

Our next example is the central limit theorem (CLT) from probability theory,
which admits a natural and well-known formulation in terms of
renormalization-group (RG) ideas
\cite{sinai1992probability,jona2001renormalization}.
Let $\{x_m\}_{m\ge1}$ be a sequence of independent and identically distributed
real-valued random variables with zero mean and unit variance.
Define the partial sums and normalized averages
\begin{equation}
S_m = x_1+\cdots+x_m, \qquad X_m = \frac{S_m}{m}.
\end{equation}
The CLT states that, as $m\to\infty$, the rescaled random variables
$\sqrt{m}\,X_m = S_m/\sqrt{m}$ converge in distribution to the standard normal
law $\mathcal{N}(0,1)$.

\subsection{Representation on the fractal lattice}

As in the previous sections, we formulate this classical result using the same
fractal space--time lattice $\mathcal{L}$ defined in
Eq.~\eqref{eq:flatt}.
We introduce a real-valued field $u_n(t)\in\mathbb{R}$ on the nodes of the
lattice.
The random variables $x_m$ are assigned to the smallest resolved
scale $n=N$, for which a single update corresponds to the turnover time
$\tau_N=2^{-N}$.
Specifically, we define
\begin{equation}
u_N(m\tau_N) = x_m, \qquad m=1,2,\ldots,
\label{eq:CLT_micro}
\end{equation}
and set $u_n(t)\equiv0$ for all $n>N$.

\begin{figure}[t]
\centering
\includegraphics[width=0.85\textwidth]{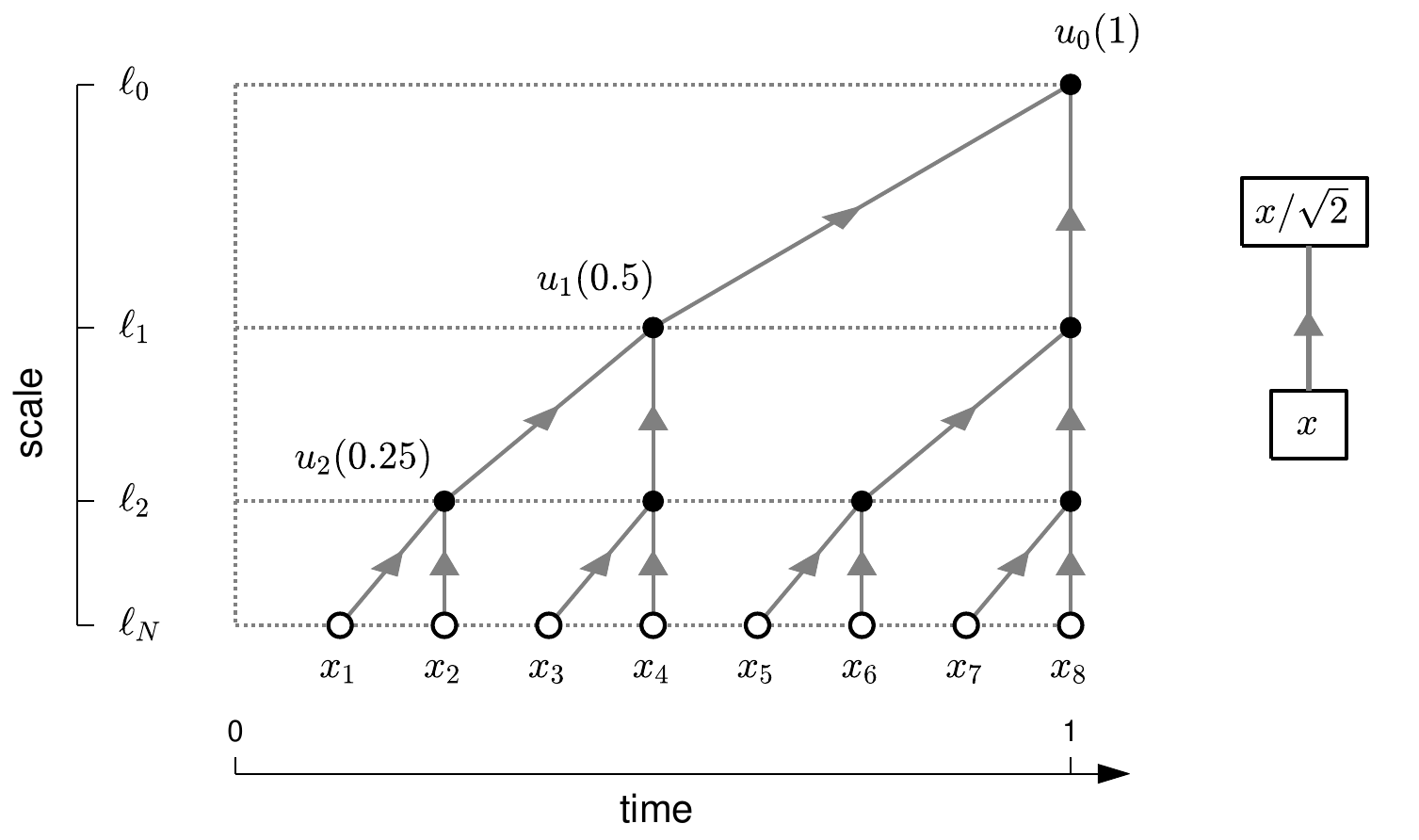}
\caption{Dynamics on the fractal lattice for the CLT.
Independent and identically distributed random variables $\{x_m\}_{m\ge0}$ are
assigned to the cutoff scale $N$.
Arrows denote the scaling relations~\eqref{eq:CLT_scaling} between the variables
$u_n(t)$ at different scales.}
\label{fig6}
\end{figure}

For scales $n<N$, the dynamics is purely self-similar and consists of
additive coupling between neighboring scales, as illustrated in
Fig.~\ref{fig6}.
Concretely, we impose the deterministic rule
\begin{equation}
u_n(t)
=
\frac{u_{n+1}(t)+u_{n+1}(t-\tau_{n+1})}{\sqrt{2}},
\label{eq:CLT_scaling}
\end{equation}
which represents the averaging of two smaller-scale contributions with
the appropriate normalization.
This prescription uniquely defines the random field $u_n(t)$ for all $n\ge0$ and
$t>0$.
In contrast to the multiscale Arnold's cat model, no initial condition is required, since the dynamics is entirely driven by the microscopic
random variables.

\subsection{RG dynamics and its fixed point}

Let $p(x)$ be a probability density of the random variables $x_m$.
This stochastic property propagates across scales, implying that each component
$u_n(t)$ admits a probability density.
Let $\varphi^{(N)}(x)$ denote the probability density of the largest-scale
variable $u_0(1)$.
In particular, for $N=0$ we have
\begin{equation}
\varphi^{(0)}(x) = p(x).
\label{eq:CLT_initial}
\end{equation}

By construction, the variable $u_0(1)$ coincides with the normalized sum
$S_m/\sqrt{m}$ for $m=2^N$.
Thus, the CLT is equivalent to the convergence of the sequence
$\{\varphi^{(N)}\}_{N \ge 0}$ to the Gaussian density
\begin{equation}
\varphi^\ast(x) = \frac{1}{\sqrt{2\pi}}\,e^{-x^2/2}.
\label{eq:CLT_gaussian}
\end{equation}

\begin{figure}[tp]
\centering
\includegraphics[width=0.6\textwidth]{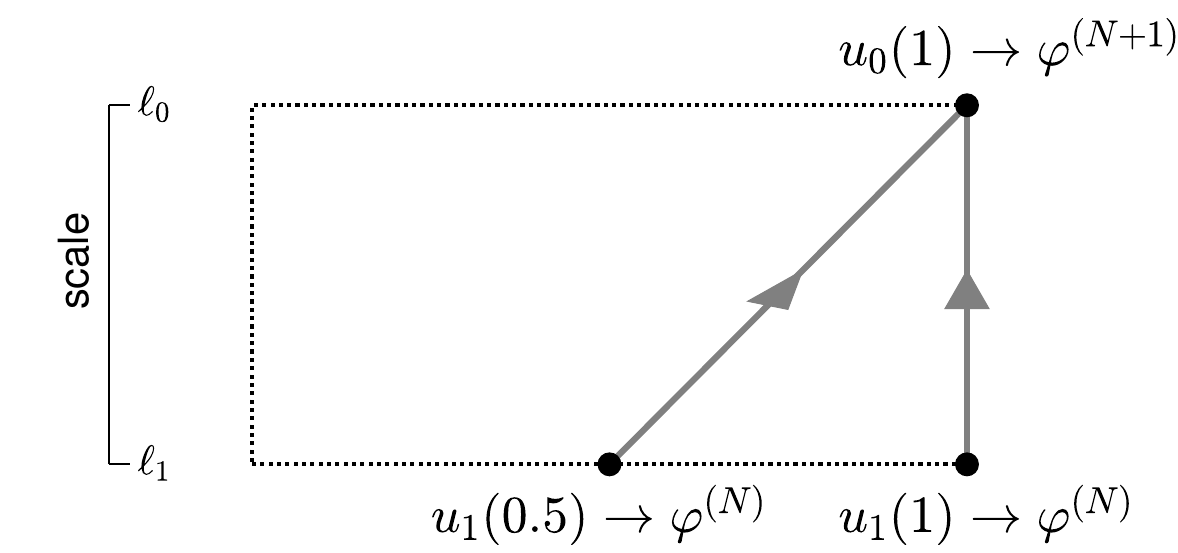}
\caption{Block structure of the cutoff-$(N+1)$ dynamics leading to the RG
equation~\eqref{eq:CLT_RG} for the central limit theorem.
At each node, we annotate the field variables and the corresponding probability
densities.
}
\label{fig7}
\end{figure}

Using the same block decomposition of the lattice as in the multiscale Arnold's
cat model, now adapted to the present setting (see Fig.~\ref{fig7}), one can relate the
densities $\varphi^{(N+1)}$ and $\varphi^{(N)}$.
Since distribution of the sum of independent random variables is given by convolution of
their probability densities, one obtains the RG relation
\begin{equation}
\varphi^{(N+1)}(x)
= \sqrt{2}
\int_{\mathbb{R}}
\varphi^{(N)}(\sqrt{2}\,x-u)\,\varphi^{(N)}(u)\,du,
\label{eq:CLT_RG}
\end{equation}
where $\sqrt{2}$ is the scaling factor from Eq.~(\ref{eq:CLT_scaling}).
This can be written compactly as an RG iteration
\begin{equation}
\varphi^{(N+1)} = \mathcal{R}[\varphi^{(N)}],
\end{equation}
where the RG operator $\mathcal{R}$ acts on probability densities according to
\begin{equation}
\mathcal{R}[\varphi](x)
= \sqrt{2}
\int_{\mathbb{R}}
\varphi(\sqrt{2}\,x-u)\,\varphi(u)\,du.
\label{eq:CLT_RG_op}
\end{equation}

As in the multiscale Arnold's cat model, the ideal limit $N\to\infty$ is thus
reduced to an RG flow in the space of probability densities, starting from the
initial condition~\eqref{eq:CLT_initial}.
The Gaussian density~\eqref{eq:CLT_gaussian} is a fixed point of the RG operator
$\mathcal{R}$ and, moreover, an attractor of the RG dynamics
\cite{sinai1992probability}.
The attracting character of this fixed point is precisely the content of the
central limit theorem.

The linearized RG equation around the Gaussian fixed point takes the form~\eqref{eq:Feig_fixed_point_lin} with
\begin{equation}
D\mathcal{R}[\psi](x)
= 
2\sqrt{2}
\int_{\mathbb{R}}
\varphi^*(\sqrt{2}\,x-u)\,\psi(u)
\, du.
\label{eq:CLT_lin}
\end{equation}
This operator can be diagonalized explicitly~\cite{sinai1992probability},
yielding eigenvalues $\rho_k = 2^{1-k/2}$ for $k = 0,1,2,\ldots$,
with eigenfunctions $\psi$ expressed in terms of Hermite polynomials.
The perturbations corresponding to the modes with $k \le 2$ are not
dynamically relevant, as they are excluded by the normalization
constraints (unit mass, zero mean, and unit variance).
Since $|\rho_k| < 1$ for all $k>2$, the Gaussian fixed point is linearly stable.
The eigenmodes with $k>2$ determine the rate at which the normalized
sums converge to the Gaussian distribution and encode the corrections
to the central limit theorem.
 
As in the previous examples, both the fixed point and its stability properties
are entirely determined by the RG operator~\eqref{eq:CLT_RG_op}, that is, by the
deterministic ideal dynamics imposed on the fractal lattice.
From this perspective, the CLT, the Feigenbaum universality, and spontaneous
stochasticity in the multiscale Arnold's cat model all arise as distinct RG
fixed-point scenarios associated with different local rules defined on the same
underlying self-similar lattice.

\section{RG for hierarchical models in statistical physics}
\label{sec7}

Hierarchical spin models provide a highly instructive setting for
the development of renormalization-group (RG) methods in statistical physics.
Originally introduced by Dyson~\cite{dyson1969existence} and further developed
by Bleher, Sinai, and others~\cite{bleher1973investigation,collet1978renormalization,jona2001renormalization},
these models feature long-range, nonlocal interactions organized in a
self-similar, hierarchical fashion.
Despite their apparent simplicity, hierarchical models capture many essential
features of critical phenomena and allow for a mathematically controlled RG
analysis.

We consider a system of $2^N$ discrete or continuous spins $x_m$, $m=1,\ldots,2^N$.
The hierarchy is introduced through the block-spin variables
\begin{equation}
S_{l,r}
=
\sum_{m=(r-1)2^l+1}^{r2^l} x_m,
\label{eq:Dyson1}
\end{equation}
defined for levels $l=0,1,\ldots,N$ and block indices
$r=1,\ldots,2^{N-l}$.
These block-spins satisfy the recursive relation
\begin{equation}
S_{l,r}
= S_{l-1,2r-1}+S_{l-1,2r}.
\label{eq:Dyson2}
\end{equation}
The interaction between spins is specified by the Hamiltonian
\begin{equation}
H_N(x_1,\ldots,x_{2^N})
=
-\sum_{l=1}^N 
\sum_{r=1}^{2^{N-l}} 
\left( \frac{\sqrt{c}}{2} \right)^{2l} \! S_{l,r}^2,
\label{eq:Dyson3}
\end{equation}
where $1<c<2$ is the interaction parameter.
This Hamiltonian satisfies the recursive decomposition
\begin{align}
H_N(x_1,\ldots,x_{2^N})
= &
\ H_{N-1}(x_1,\ldots,x_{2^{N-1}}) \\
& + H_{N-1}(x_{2^{N-1}+1},\ldots,x_{2^N})
- \left( \frac{\sqrt{c}}{2} \right)^{2N} \!  S_{N,1}^2,
\label{eq:Dyson4}
\end{align}
which makes explicit the hierarchical structure of the nonlocal interaction.

The Gibbs measure of the model is defined by
\begin{equation}
d\mu_N(x_1,\ldots,x_{2^N})
=
Z_N^{-1}
e^{-\beta H_N(x_1,\ldots,x_{2^N})}
\prod_{m=1}^{2^N} d\mu_0(x_m),
\label{eq:Dyson5}
\end{equation}
where $\beta$ is the inverse temperature, $d\mu_0(x)$ is a given single-spin distribution, and
$Z_N$ is the normalization constant.

\subsection{Representation on the fractal lattice}

\begin{figure}[t]
\centering
\includegraphics[width=0.85\textwidth]{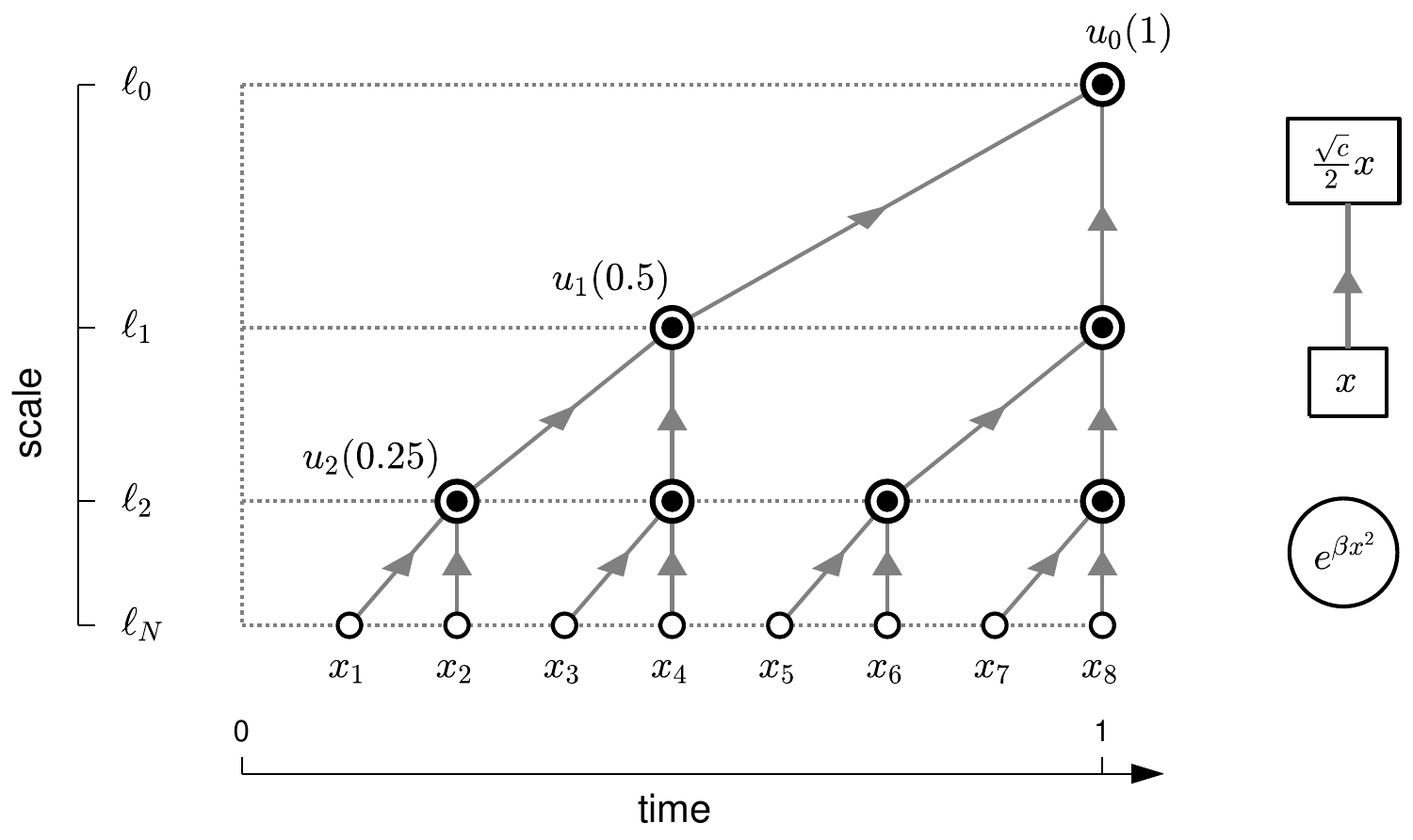}
\caption{
Fractal-lattice representation of the hierarchical spin model.
Independent and identically distributed random variables
$\{x_m\}_{m\ge 1}$ are assigned at the cutoff scale $N$.
Arrows indicate the scaling relations~\eqref{eq:Dyson6}
between the variables $u_n(t)$ across scales.
Large circles represent the local self-interaction factors~\eqref{eq:Dyson6b}.
}
\label{fig8}
\end{figure}

Let us show how the hierarchical model can be represented as dynamics on the fractal lattice $\mathcal{L}$; see Fig.~\ref{fig8}. We introduce real-valued random variables $u_n(t)$ defined at the lattice nodes, with a cutoff at scale $n = N$. The cutoff variables $u_N(t) = x_m$ at discrete times $t = m\tau_N$ are assumed to be independent and identically distributed with law $d\mu_0(x_m)$, corresponding to single non-interacting spins. For $n > N$, we set $u_n(t) \equiv 0$. It remains to establish relations among the variables $u_n(t)$ at scales $n < N$.

Unlike the previous examples, where the dynamical relations were entirely deterministic, the present relations contain both deterministic and stochastic components.
In the deterministic part, we define
\begin{equation}
\tilde{u}_n(t) =
\frac{\sqrt{c}}{2}
\big[u_{n+1}(t)+u_{n+1}(t-\tau_{n+1})\big].
\label{eq:Dyson6}
\end{equation}
This relation is illustrated by arrows in Fig.~\ref{fig8}. 
The distribution of $u_n(t)$ is then obtained by reweighting the probability measure $d\mu(x)$ of the random variable $x = \tilde{u}_n(t)$ according to
\begin{equation}
d\mu(x) \ \mapsto \
\frac{1}{Z} \, e^{\beta x^2} d\mu(x),
\label{eq:Dyson6b}
\end{equation}
where $Z$ ensures normalization to unit mass.
This reweighting step is indicated by the large circles in Fig.~\ref{fig8}.

The lattice relations are constructed so as to reproduce the hierarchical
structure~\eqref{eq:Dyson4} of the Gibbs distributions~\eqref{eq:Dyson5}.
Indeed, one verifies that each lattice variable $u_n(t)$ for $n \le N$
admits a representation in terms of properly normalized block spins~\eqref{eq:Dyson1} as
\begin{equation}
u_n(t)
=
\left( \frac{\sqrt{c}}{2} \right)^{l}
\sum_{m=(r-1)2^l+1}^{r2^l} x_m,
\qquad
l = N-n,
\label{eq:Dyson_S1}
\end{equation}
where the joint distribution
$\mu_l\big(x_{(r-1)2^l+1},\ldots,x_{r2^l}\big)$ is the Gibbs measure incorporating all interactions 
restricted to the corresponding block.

\subsection{RG equation and critical behavior}

Our quantity of interest is the macroscopic variable
\[
u_0(1) = \left( \frac{\sqrt{c}}{2} \right)^{N}
\sum_{m=1}^{2^N} x_m,
\]
whose distribution encodes the thermodynamic properties of the full Gibbs measure~(\ref{eq:Dyson5}).
Let $\varphi^{(N)}(x)$ denote the corresponding probability density of
$x = u_0(1)$.
By construction, the single-spin distribution defines
\begin{equation}
\varphi^{(0)}(x) = \mu_0(x),
\label{eq:Dyson7}
\end{equation}
which serves as the initial condition for the RG flow.

Similarly to the CLT example, see Eq.~(\ref{eq:CLT_RG}) and Fig.~\ref{fig7},
relation~(\ref{eq:Dyson6}) yields a rescaled convolution relation for the
corresponding distributions.
Combining this with the additional interaction term~(\ref{eq:Dyson6b}),
one obtains the recursion relation~\cite{bleher1973investigation,jona2001renormalization}
\begin{equation}
\varphi^{(N+1)}(x)
=
z_N\,
e^{\beta x^2}
\int_{\mathbb{R}}
\varphi^{(N)}\!\left(\frac{2x}{\sqrt{c}}-u\right)
\varphi^{(N)}(u)\,du,
\label{eq:Dyson9}
\end{equation}
where $z_N$ is a normalization constant.
This recursion defines an RG iteration
\begin{equation}
\varphi^{(N+1)} = \mathcal{R}[\varphi^{(N)}],
\label{eq:Dyson10}
\end{equation}
with RG operator
\begin{equation}
\mathcal{R}[\varphi](x)
=
z\,
e^{\beta x^2}
\int_{\mathbb{R}}
\varphi\!\left(\frac{2x}{\sqrt{c}}-u\right)
\varphi(u)\,du,
\label{eq:Dyson11}
\end{equation}
where $z$ normalizes the density to unit mass.
The thermodynamic limit $N \to \infty$ is thus represented as an RG dynamics in the
space of probability distributions.

The RG operator $\mathcal{R}$ admits a Gaussian fixed point~\cite{bleher1973investigation,jona2001renormalization}
\begin{equation}
\varphi^*(x)
=
\frac{1}{\sqrt{2\pi\sigma^2}}
e^{-x^2/(2\sigma^2)},
\quad
\sigma^2 = \frac{2-c}{2c\beta},
\label{eq:Dyson12}
\end{equation}
whose variance depends explicitly on the interaction parameter $c$ and the
inverse temperature $\beta$.
A detailed stability analysis~\cite{bleher1973investigation,jona2001renormalization}
shows that this fixed point possesses a single relevant (unstable) direction for
$\sqrt{2}<c<2$, and therefore governs the critical behavior of the model in this
parameter range.
At $c=\sqrt{2}$, a bifurcation occurs in which a second unstable direction
appears.
For $1<c<\sqrt{2}$, this leads to the emergence of a non-Gaussian RG fixed point,
which then controls the phase transition in this regime.

Once again, we see that the macroscopic universality class is fully determined
by the fixed points and stability properties of an RG operator defined on the
same underlying fractal lattice.
From this perspective, the RG formulation of hierarchical models naturally
fits alongside the previously discussed examples from turbulence, dynamical
systems, and probability theory, all of which can be viewed as distinct
realizations of self-similar dynamics on a common multiscale structure.

\section{Discussion and synthesis}

In this work, we have presented a unified renormalization-group (RG) perspective
on spontaneous stochasticity and several classical universality phenomena,
including the Feigenbaum period-doubling cascade, the central limit theorem, and
hierarchical spin models of statistical physics.
Although these theories originate in different physical and mathematical
contexts, we have shown that they can all be formulated within a common
multiscale framework based on dynamics defined on the same self-similar
(fractal) space--time lattice.
The essential distinction between the examples lies in the local dynamical rules, 
which determine the form of RG operators.

Across all cases considered, the macroscopic limit is reduced to the iteration
of an RG transformation acting on an effective description of the dynamics:
deterministic maps in the Feigenbaum case, probability densities in the CLT and
hierarchical spin models, and Markov kernels in the multiscale Arnold's cat
model.
Universality emerges when this RG dynamics converges to a fixed
point, whose properties depend only on the RG operator and not on the
microscopic details encoded in the initial condition.
In this sense, universality is a manifestation of the loss of microscopic
information under successive RG transformations.

The multiscale Arnold's cat model provides a particularly transparent
illustration of this mechanism in a dynamical context.
In this model, the ideal system admits uncountably many deterministic
solutions, reflecting a fundamental loss of well-posedness caused by
the propagation of information from arbitrarily small to large scales.
Introducing a cutoff together with microscopic noise yields a unique
stochastic evolution at each finite cutoff, which converges to a
universal inviscid limit as the cutoff is removed.
From the RG perspective, spontaneous stochasticity corresponds to
convergence toward a nontrivial fixed point of the RG dynamics acting
on Markov kernels.
This fixed point provides a canonical probabilistic resolution of
non-uniqueness, dynamically selected and independent of the specific
regularization details.

Seen from this angle, spontaneous stochasticity fits naturally alongside more
classical RG phenomena.
In the CLT, the Gaussian distribution plays the role of a stable RG fixed point
for the addition-and-rescaling operator.
In the Feigenbaum scenario, the period-doubling universality is governed by a
nontrivial RG fixed point with a one-dimensional unstable manifold.
In hierarchical spin models, Gaussian and non-Gaussian fixed points control the
critical behavior depending on model parameters.
In all these examples, the macroscopic laws are encoded in the fixed points and
stability properties of an RG operator defined on the same multiscale lattice.

At the same time, the present framework also highlights important limitations
and open challenges.
All models considered here involve either deterministic microscopic dynamics or
noise that is independent and identically distributed at the smallest resolved
scale.
This assumption greatly simplifies the RG analysis and ensures that successive
coarse-graining steps produce dynamics with controlled statistical
properties.
In more realistic models of statistical physics and fluid dynamics, however,
small-scale degrees of freedom may be correlated in both space
and time.
As is well known from statistical physics, such correlations may persist
across scales and influence the macroscopic state.
Another major challenge concerns dimensionality.
The fractal lattice employed here is effectively one-dimensional 
(or even zero-dimensional in the context of fluid dynamics), while realistic systems—such as turbulent flows or lattice spin
models—live in higher-dimensional physical spaces.
Extending the present RG framework to incorporate multi-dimensional interactions and correlations
remains a nontrivial task.

Despite these limitations, we believe that the present work clarifies the
conceptual role of spontaneous stochasticity within the broader RG paradigm.
Rather than being a pathological feature of singular limits, spontaneous
stochasticity emerges here as a natural and robust universality phenomenon,
associated with RG fixed points in spaces of probability measures or kernels.
This perspective suggests that probabilistic descriptions of ideal limits are
not merely unavoidable, but in fact intrinsic to multiscale systems with strong
information transfer across scales.

We hope that this unified perspective will foster connections between
turbulence, dynamical systems, probability theory, and statistical
physics, and inspire further development of renormalization-group
approaches to stochasticity and non-uniqueness in more realistic models.

\begin{acknowledgement}
A.A.M. is grateful to Theodore D. Drivas for valuable discussions on RG methods in probability theory and to Nazarbayev University for supporting his visit to Astana. This work was also supported by CNPq grants No.~308721/2021-7 and 311012/2022-1, and by the CAPES MATH-AmSud project CHA2MAN.
\end{acknowledgement}

\ethics{Competing Interests}{The authors have no conflicts of interest to declare that are relevant to the content of this chapter.}

\bibliographystyle{plain}
\bibliography{refs}

\end{document}